\documentclass[11pt,a4paper]{article}
\usepackage{jheppub}
\usepackage{subfigure}
\usepackage{feynmf}
\input epsf

\def\Ansatz{\textit{Ansatz}}
\def\Eq#1{Eq.~(\ref{#1})}
\def\Abs#1{\left\vert{#1}\right\vert}
\def\be{\begin{equation}}
\def\ee{\end{equation}}
\def\bea{\begin{eqnarray}}
\def\eea{\end{eqnarray}}
\def\nn{\nonumber}
\def\Fg{{\boldsymbol{F}}_{\!\!g}}

\def\ptwmax{\tilde{p}_{\rm max}}
\def\pmax{p_{\rm max}}
\def\barf{\bar{f}}
\def\barT{\bar{T}_*}

\def\calC{{\mathcal{C}}}
\def\tcalC{\tilde{\mathcal{C}}}

\def\nc{N_{\rm c}}

\def\twotwo{{2 \leftrightarrow 2}}
\def\twothree{{2 \leftrightarrow 3}}
\def\onetwo{{1 \leftrightarrow 2}}
\def\OO{{\cal O}}
\def\ptw{{\tilde{p}}}
\def\qtw{\tilde{q}}
\def\ktw{{\tilde{k}}}
\def\ftw{\tilde{f}}
\def\ntw{\tilde{n}}
\def\gammatw{\tilde{\gamma}}

\def\lsim{\mbox{~{\raisebox{0.4ex}{$<$}}\hspace{-1.1em}
        {\raisebox{-0.6ex}{$\sim$}}~}}

\newcommand{\V}{\boldsymbol}

\newcommand{\mD}{m_{_{\rm D}}}
\newcommand{\mDsq}{\mD^2}
\newcommand{\mDtw}{\tilde{m}_{_{\rm{D}}}}
\newcommand{\mDtwsq}{\mDtw^2}

\preprint{CERN-PH-TH/2014-006}
\title{UV Cascade in Classical Yang-Mills via Kinetic Theory}
\author[a]{Mark C. Abraao York,}
\author[a,b]{Aleksi Kurkela}
\author[a]{Egang Lu}
\author[a]{Guy D. Moore}
\affiliation[a]{McGill University Department of Physics\\
3600 Rue University \\ Montr\'eal, QC H3A 2T8, Canada}
\affiliation[b]{CERN, Physics Department, 1211 Gen\`eve 23, Switzerland}

\abstract{
We show that classical Yang-Mills theory with statistically homogeneous
and isotropic initial conditions has a kinetic description and
approaches a scaling solution at late times.  We find the scaling
solution by explicitly solving the Boltzmann equations, including
all dominant processes (elastic and number-changing).  Above a
scale $\ptwmax \propto t^{\frac 17}$ the occupancy falls
exponentially in $p$.
For asymptotically late times and sufficiently small momenta
the occupancy scales as $f(p)\propto 1/p$,
but this behavior sets in only at very late time scales.
We find quantitative agreement of our results with
lattice simulations, for times and momenta within the range of validity of
kinetic theory.
}

\begin{document}

\maketitle

\section{Introduction}

Recently there have been a number of studies of classical Yang-Mills
theory, motivated by the expectation that it describes quantum
Yang-Mills theory in the dual limits of weak coupling and high
occupancy.  The study of this limit is motivated by arguments that the
conditions early after a heavy ion collision are described by the
``glasma,'' which is precisely such a weak-coupled but high-occupancy
state \cite{Gelis,Lappi,Weigert}.  Recent theoretical attempts to
describe the dynamics of classical Yang-Mills theory
\cite{KM1,BGLMV} already differ in some details when describing the
simplest case of statistically homogeneous and isotropic, non-expanding
initial conditions, which has motivated numerical lattice studies of
this limit \cite{Berges,KM3,Schlichting}.

These studies show that, as expected, classical Yang-Mills theory has no
equilibrium state, but features a self-similar cascade of energy from
the infrared towards the ultraviolet, with the typical momentum of an
excitation rising with time as $t^{\frac 17}$ and the typical occupancy
decaying as $t^{\frac{-4}7}$
\cite{Berges,KM3,Schlichting}.  These lattice studies
suffer from a limited dynamic and temporal range, statistical errors
particularly in the infrared, and lattice spacing corrections.  On the
other hand, at late times the dynamics should also be well described by
kinetic theory.  A kinetic study also allows the possibility to better
investigate the details of the cascade.  For instance, in the kinetic
description we can better determine the relative importance of elastic
versus inelastic processes, and whether the latter are efficient at all
energies or only in the infrared.

Therefore in the current paper we will revisit the problem of the
cascade to the ultraviolet in classical Yang-Mills theory, using kinetic
theory.  In the next section we review the problem, the scales involved,
and the form of kinetic theory (with some details about kinetic theory
postponed to an appendix).

In Section \ref{secSolving} we introduce two approaches to treating
kinetic theory numerically. In Section \ref{results}, we find that within the domain of its applicability the kinetic theory reproduces the lattice simulations with great accuracy (see Figure \ref{mainresult}), but
the treatment is numerically far less demanding. The increased accuracy then
allows us to study the scaling solution in far greater detail than on the lattice. In particular, we present evidence that the scaling solution scales as $f(p)\propto 1/p$ (Figure \ref{md_scaling}). Finally we conclude with a Summary.

\section{Boltzmann equation and scaling solution in classical Yang-Mills}

Consider quantum Yang-Mills theory where the (t'Hooft) coupling is weak
$g^2 \nc \ll 1$, while the mean occupancy is high, $f \gg 1$.  In this
regime the
theory is well described by the classical approximation.  If in addition
$g^2\nc f \ll 1$ then the classical theory is in a weakly coupled regime
and kinetic theory should be applicable (see e.g.~\cite{Mueller:2002gd}).
Since $g^2 N_c f$ controls the
weak-coupling expansion, we will introduce $\bar{f}=g^2 \nc f$;
weak coupling is $\barf \ll 1$.

Classical Yang-Mills theory has no equilibrium;
if we start off initially with a system where the energy
density resides below some scale $\pmax$, this scale will grow
with time.   In fact, at late times we expect the occupancy
to evolve towards a scaling solution \cite{KM3}. To see this, we first
introduce a characteristic energy scale
$Q$, determined by the energy density $\mathcal{E}$ via
\be
Q^4 = \frac{\pi^2 g^2 \nc \mathcal{E}}{\nc^2 - 1} , \qquad \mbox{so}
\qquad Q^4 = \int p^3 \barf(p) \: dp \quad \mbox{in kinetic theory.}
\label{Qdef}
\ee
Next we examine how $\barf(p)$ evolves with time under the Boltzmann
equation.  For the moment we consider $\twotwo$ scatterings;
\bea
\label{Boltzmann0}
\frac{\partial f(p,t)}{\partial t} & = & - \calC_{\twotwo}[f(p,t)] \,,
\\
\calC_{\twotwo}[f_p] &=&  \frac{1}{2\nu_g}
  \int \frac{d^3k}{(2\pi^3)}\frac{d^3p'}{(2\pi^3)}\frac{d^3k'}{(2\pi^3)} \frac{\Abs{\overline{{\cal M}}_{pk;p'k'}^2}}{2p2k2p'2k'} \;
  (2\pi)^4 \delta^4(p{+}k{-}p'{-}k') \times
\nn \\ &&\hspace{1.4cm}
  \Big( f_p f_k[1+f_{p'}][1+f_{k'}] -[1+f_p][1+f_k]f_{p'}f_{k'} \Big) \,.
\nn
\eea
Here $k$ is the other incoming, and $p',k'$ the outgoing, momenta,
$\Abs{\overline{{\cal M}}_{pk;p'k'}^2}$ is the squared matrix element summed
(not averaged) over all external colors and spins, and the last line is
the difference of the statistical factors for the processes with $p$ as
an initial state (first term) and the inverse process with $p$ as a
final state (second term). The number of degrees of freedom is denoted by $\nu_g$, which for gauge bosons reads $2 d_A = 2 (N_c^2-1)$.
Since we consider $f\gg 1$, we may simplify the occupancies,
\bea
\hspace{-1ex}
\Big( f_p f_k[1+f_{p'}][1+f_{k'}]
 - [1+f_p][1+f_k]f_{p'}f_{k'} \Big)
& \simeq & f_p f_k f_{k'} + f_p f_k f_{p'} - f_k f_{p'} f_{k'}
           - f_p f_{p'} f_{k'} \quad
\nn \\ & = & f_p f_k f_{p'} f_{k'}
    \Big( f_{p'}^{-1} + f_{k'}^{-1} - f_p^{-1} - f_k^{-1} \Big) .
\eea
This simplification amounts to making the classical field approximation.

The external state summed, squared matrix element
$\Abs{\overline{\cal M}}^2$ naturally scales as $g^4 \nc^2 \nu_g$,
\be
\label{M22}
\vert \overline{\mathcal{M}}^2_{pk;p'k'} \vert = 2 g^4 \nc^2 \nu_g \Big ( 9 + \frac{(t-u)^2}{s^2}+\frac{(s-u)^2}{t^2}+ \frac{(s-t)^2}{u^2}\Big ) .
\ee
Defining
$\Abs{\overline{M}^2_{pk;p'k'}} = \Abs{\overline{\cal M}}^2/(g^4 \nc^2 \nu_g)$, we
find that factors of $g^2 \nc$ cancel on both sides when we rewrite
\Eq{Boltzmann0} in terms of $\barf$ and $\Abs{\overline{M}}^2$;
\bea
\label{Boltzmann1}
\frac{d \barf(p,t)}{d t} & = & -\frac{1}{2}
  \int \frac{d^3 k}{(2\pi^3)}\frac{d^3p'}{(2\pi^3)}
 \frac{d^3k'}{(2\pi^3)} \frac{\Abs{\overline{M}_{pk;p'k'}^2}}{2p2k2p'2k'} \;
  (2\pi)^4 \delta^4(p{+}k{-}p'{-}k') \times
\nn \\ &&\hspace{1.4cm}
 \barf_p \barf_k \barf_{p'} \barf_{k'}
    \Big( \barf_{p'}^{-1} + \barf_{k'}^{-1}
    - \barf_p^{-1} - \barf_k^{-1} \Big) \,.
\eea

Now assume that the characteristic momentum scale grows as a fractional
power of time, $p \propto t^{\alpha}$.  By \Eq{Qdef} and energy
conservation, the typical occupancy will then fall,
$\barf \propto t^{-4\alpha}$.  So we introduce dimensionless momentum
and occupancy variables which account for this scaling behavior;
\bea
\label{ptw}
\ptw(p,t) & \equiv & (p/Q) (Qt)^{-\alpha} \qquad \mbox{so} \quad
p = \ptw \, Q (Qt)^\alpha \,, \\
\label{ftw}
\ftw(\ptw,t) & \equiv & (Qt)^{4\alpha} \barf(p,t)
\qquad \mbox{so} \quad
\barf(p,t) = \ftw(\ptw,t) (Qt)^{-4\alpha} \,.
\eea
In terms of these variables, the lefthand side of the Boltzmann equation
becomes
\be
\label{twotwoscales}
\frac{d \barf(p,t)}{d t} =
   \frac{d [ (Qt)^{-4\alpha} \ftw( (Qt)^{\alpha} Q \ptw,t)]}{dt}
 = (Qt)^{-4\alpha} \;
 \frac{\alpha}{t} \left( -4 \ftw(\ptw)
                   - \ptw \frac{\partial \ftw(\ptw,t)}{\partial \ptw}
                   + \frac{t}{\alpha}\frac{\partial \ftw(\ptw,t)}{\partial t}
 \right)  \,,
\ee
where the last term is the explicit $t$ dependence of $\ftw$, that is,
the time dependence not incorporated into the time scaling we have
applied.  The righthand, collision side of \Eq{Boltzmann1} involves 1
power of momentum since $\Abs{\overline{M}}^2$ is dimensionless, and it
contains three powers of $\barf$; so rescaling in terms of
$\ptw$ and $\ftw$ scales out a factor of
$Q (Qt)^\alpha (Qt)^{-12\alpha}= (Qt)^{1-11\alpha}/t$.
In order for the left and right hand sides to scale in the same way with
$(Qt)$, we must therefore have
\be
(Qt)^{-4\alpha} t^{-1} = (Qt)^{1-11\alpha} t^{-1} \qquad \mbox{or}
\qquad
\alpha = 1/7 \,.
\ee
This reproduces the time scaling behavior found in \cite{KM1,BGLMV}.

Therefore, the Boltzmann equation becomes
\be
\frac{t \partial \ftw(\ptw,t)}{\partial t} =
\frac{1}{7} \left( 4 \ftw(\ptw)
                   + \ptw \frac{d\ftw(\ptw)}{d\ptw} \right)
 -\tcalC_{\twotwo}[\ftw(\ptw)] \,.
\label{lhs=rhs}
\ee
Here $\tcalC_\twotwo$ is the righthand side of \Eq{Boltzmann1} but with
$p,\barf \rightarrow \ptw,\ftw$.  This expression is a relaxation
equation for $\ftw$ to approach a ``scaling'' form where it possesses no
explicit time dependence, so the two terms on the righthand side of
\Eq{lhs=rhs} cancel.  We expect $\ftw$ to approach this scaling form
(tracking solution) rather quickly -- an expectation supported by
lattice studies \cite{KM3} -- so we will focus on determining the
scaling solution itself.

Higher-order scattering processes, that is, those with more
participating external lines, are naively suppressed.  For instance, a
$\twothree$ process would involve an extra vertex, an extra momentum
integration, and an extra external state statistical factor.  The vertex
and statistical factor give rise to a factor of $g^2 \nc f = \barf$.
The two powers of momentum in the integration measure $d^3 l/l$ are
canceled by the matrix element becoming dimensionful,
$\Abs{M}^2 \sim 1/p^2$.  So {\em at generic energies and angles}, the
$\twothree$ process is suppressed, relative to the $\twotwo$ process, by
a factor of $\barf \sim (Qt)^{-4/7}$.  This is why we previously stated
that $\barf \ll 1$ is the criterion for perturbative, kinetic behavior.

There are exceptions to this argument, when the matrix element
possesses sufficiently strong soft and/or collinear divergences.  When
such divergences occur, it is necessary to include screening effects to
produce finite and correct expressions for the scattering term.  In
nonabelian gauge theory this is actually already necessary for the
$\twotwo$ process we have been discussing; $\Abs{\overline M}^2$ is
quadratically divergent in the $(\ptw-\ptw')\rightarrow 0$ limit, giving
rise to a log divergence in \Eq{lhs=rhs} (only logarithmic because
$\ftw_p^{-1}-\ftw_{p'}^{-1}$ nearly cancels in this limit).  To handle
this divergence correctly, we must incorporate screening effects (Hard
Loops) in the computation of $\Abs{\overline M}^2$.  The technical
complications have been considered elsewhere \cite{AMY5}; we will
discuss them a little more in Section \ref{secSolving}.  Here we just
remark that the would-be log divergence is regulated by the scale
$\mD$, which is parametrically
\be
\label{mD_is}
\mDsq = 4 \int \frac{d^3 p}{(2\pi)^3} \frac{\barf(p)}{p}
 \sim Q^2 (Qt)^{-2/7} \quad \mbox{so} \quad
\frac{\mD}{\pmax} \sim \frac{Q (Qt)^{-1/7}}{Q (Qt)^{+1/7}}
 \sim (Qt)^{-2/7} \,.
\ee
Therefore the scale which regulates infrared effects in the collision
term actually changes gradually with time.  To keep track of this
change, we introduce
\be
\mDtw \equiv \frac{\mD}{Q (Qt)^{1/7}} \quad ( \sim (Qt)^{-2/7} )
\ee
which keeps track of $\mD$ in the same dimensionless units as we use for
momenta $\ptw$.  Because $\mDtw$ varies with time, our argument for a
scaling solution is not quite correct.  Because the
time dependence of $\mDtw$ is very weak we expect this to be a minor
effect and we will still seek a scaling solution.

The Debye scale also plays the role of the infrared scale beyond which
the kinetic theory description is no longer reliable.  This is
because our kinetic description assumes that the dispersion relation is
lightlike and the spectral function carries all its weight on a
quasiparticle pole, properties which break down at this scale.
Therefore our results are not to be trusted at and below the scale
$\ptw = \mDtw$.

We saw above that, at generic momenta and angles, higher-leg processes
are suppressed.  But they are unsuppressed in any soft or collinear
phase space region where they are sufficiently
soft and collinear divergent.  We see from the above arguments that to
be relevant at late times, a
$\twothree$ process must be quadratically soft and/or collinear
divergent to introduce a factor of $\ptw^2/\mDtwsq \sim (Qt)^{4/7}$ which
compensates the factor $\barf\sim (Qt)^{-4/7}$.  Additional lines
require stronger power divergences.  Arnold, Moore, and Yaffe showed
that sufficiently strong divergences occur only in $n\leftrightarrow
(n+1)$ processes; and that such processes can be treated in terms of an
effective $\onetwo$ process \cite{AMY5}.  We show in Appendix
\ref{secM12} that in the current context these $\onetwo$ processes scale
with time in the same way as $\twotwo$ processes and must be included in
the collision term.  Explicitly, \Eq{lhs=rhs} becomes
\be
\label{Boltzmann2}
0 = \frac{1}{7} \left( 4 \ftw(\ptw)
                       + \ptw \frac{d\ftw(\ptw)}{d\ptw} \right)
 - \tcalC_\twotwo [ \ftw(\ptw)]
 - \tcalC_\onetwo [ \ftw(\ptw)] \,,
\ee
where
\bea
\label{C2x2tw}
\tcalC_\twotwo [ \ftw(\ptw)] & = & \frac{1}{2}
  \int \frac{d^3 \ktw}{(2\pi^3)}\frac{d^3\ptw'}{(2\pi^3)}
 \frac{d^3\ktw'}{(2\pi^3)} \frac{\Abs{\overline{M}_{\ptw \ktw;\ptw'\ktw'}^2}}{2\ptw2 \ktw 2\ptw'2\ktw'} \;
  (2\pi)^4 \delta^4(\ptw{+}\ktw{-}\ptw'{-}\ktw') \times
\nn \\ &&\hspace{1.4cm}
 \ftw_{\ptw} \ftw_{\ktw} \ftw_{\ptw'} \ftw_{\ktw'}
    \Big( \ftw_{\ptw'}^{-1} + \ftw_{\ktw'}^{-1}
    - \ftw_{\ptw}^{-1} - \ftw_{\ktw}^{-1} \Big)
\eea
and
\bea
\label{C1x2tw}
\tcalC_\onetwo [\ftw(\ptw)] & = & \frac{(2\pi)^3}{\ptw^2}
 \int_0^{\frac{\ptw}{2}} d\ktw \: \gammatw^g_{gg}(\ptw;\ktw,\ptw{-}\ktw)
 \Big( \ftw_\ptw \ftw_{\ptw{-}\ktw} + \ftw_\ptw \ftw_\ktw
      - \ftw_\ktw \ftw_{\ptw{-}\ktw} \Big)
\nonumber \\ && +  \frac{(2\pi)^3}{\ptw^2}
 \int_0^{\infty} d\ktw \: \gammatw^g_{gg}(\ptw{+}\ktw;\ptw,\ktw)
 \Big( \ftw_\ptw \ftw_\ktw - \ftw_\ptw \ftw_{\ptw{+}\ktw}
      - \ftw_\ktw \ftw_{\ptw{+}\ktw} \Big)
\,;
\eea
this term is explained and the splitting rate $\gammatw^g_{gg}$
is defined in Appendix \ref{secM12}.

\section{IR and UV limiting behaviors}

Before solving \Eq{Boltzmann2}, it is useful to study analytically how
the solution should scale with $\ptw$ in the IR and in the UV.  Knowing the
scaling behavior will also be helpful when we attempt a numerical
solution.

We begin with the UV limiting behavior.  We expect
the large $\ptw$ behavior of $\ftw(\ptw)$ to be exponential,
$\ftw(\ptw) \rightarrow \ptw^a \exp(- b \ptw)$ for some $a,b$.  To see
this, note first that, for the energy to be bounded, the occupancy in
the tail has to fall faster than $\ftw \propto \ptw^{-4}$.  But
in any region where $|\ptw d\ftw/d\ptw| > 4 \ftw$,
\Eq{twotwoscales} shows that $d\barf/dt > 0$ -- scatterings must move
particles {\sl into} the UV tail, at a rate comparable to the system
age.  Exponential behavior is
self-consistent, because to produce a particle with $\ptw \gg 1$ one
must scatter or merge together particles of energies totalling at
least $\ptw$; exponential behavior means that the final state occupancy
scales with the likelihood of finding two constituents which are
available to merge.
Stimulation factors do not change this argument.
Super-exponential behavior such as $\ftw \sim \exp(-k \ptw^2)$ can be
excluded, because there are far more pairs of particles of energy
$\ptw/2$ available than particles of energy $\ptw$; so the merger rate
to momentum scale $\ptw$ would greatly exceed the occupancy there.
Similarly, power-law UV behavior cannot provide enough scatterings to
keep the tail growing.

To explore the infrared behavior of $\ftw(\ptw)$, it is useful to
consider total particle number.  By integrating the Boltzmann equation,
\Eq{Boltzmann2}, over momentum $\int \frac{d^3 \ptw}{(2\pi)^3}$, we obtain
an equation describing total particle number change;
\be
\int \frac{\ptw^2 d\ptw}{2\pi^2} \left(
\left[ \frac{4}{7} \ftw_\ptw + \frac{\ptw d\ftw_\ptw}{7 d\ptw} \right]
- \tcalC_{\onetwo}[\ftw_\ptw]   - \tcalC_{\twotwo}[\ftw_\ptw] \right) = 0 \,.
\label{partnum1}
\ee
The contribution from the term in square brackets is
\be
\frac{1}{14\pi^2} \int \ptw^2 \Big( 4 \ftw_\ptw + \ptw \ftw'_\ptw \Big)
d\ptw = \frac{1}{14 \pi^2} \int \ptw^2 \ftw_\ptw \: d\ptw
 = \frac{\ntw}{7}
\label{partnumL}
\ee
which (up to our rescalings by factors of $g^2 \nc$ and $(Qt)$) is just
minus the time rate of change of particle number,
$-t dn/dt = n/7$, which is finite.  The contribution from
$\twotwo$ processes is, unsurprisingly, zero;
\bea
\int \frac{d^3\ptw}{(2\pi)^3} \: \tcalC_{\twotwo}[\ftw_\ptw]
& = & \int \frac{d^3 \ptw\, d^3 \ktw \,d^3 \ptw'\, d^3 \ktw'}{(2\pi)^9}
 \frac{\Abs{ \overline{M}^2_{\ptw \ktw;\ptw'\ktw'}}^2}
      {16 \ptw \ktw \ptw' \ktw'}
      (2\pi)^4 \delta^4(\ptw {+} \ktw {-} \ptw' {-} \ktw' )
\nonumber \\ & & \hspace{2.4cm} \times
    \ftw_\ptw \ftw_\ktw \ftw_{\ptw'} \ftw_{\ktw'}
   \Big( \ftw_{\ptw'}^{-1} + \ftw_{\ktw'}^{-1}
         - \ftw_{\ptw}^{-1} - \ftw_{\ktw}^{-1} \Big)
\label{partnum22}
\eea
which vanishes since the first line is symmetric, and the second
antisymmetric, on exchanging primed and unprimed variables.
The contribution from $\onetwo$ processes is
$\int \frac{d^3 \ptw}{(2\pi)^3}$ of \Eq{C1x2tw}, which is
\bea
\label{partnum12_1}
\int \frac{d^3\ptw}{(2\pi)^3} \tcalC_{\onetwo}[\ftw_\ptw]
 & = & \phantom{+}4\pi \int_0^\infty d\ptw \int_0^{\ptw/2} d\ktw\:
  \gammatw^g_{gg}(\ptw;\ktw,\ptw-\ktw)
 \Big( \ftw_\ptw \ftw_{\ptw-\ktw} + \ftw_\ptw \ftw_{\ktw}
  - \ftw_{\ktw} \ftw_{\ptw-\ktw} \Big)
 \nonumber \\ && + 4\pi \int_0^\infty d\ptw d\ktw\:
  \gammatw^g_{gg}(\ptw+\ktw;\ktw,\ptw)
 \Big( \ftw_\ptw \ftw_{\ktw} - \ftw_{\ptw+\ktw} \ftw_{\ktw}
  - \ftw_{\ptw+\ktw} \ftw_{\ptw} \Big) \,.
\eea
To make the two terms look more similar we rename $\ptw$ in the first
equation to $\ptw{+}\ktw$;
\bea
\label{partnum12_2}
\int \frac{d^3\ptw}{(2\pi)^3} \tcalC_{\onetwo}[\ftw_\ptw]
& = & \phantom{+}4\pi \int_{\ptw>\ktw} d\ptw d\ktw \;
  \gammatw^g_{gg}(\ptw+\ktw;\ktw,\ptw)
 \Big( \ftw_{\ptw+\ktw} \ftw_{\ptw} + \ftw_{\ptw+\ktw} \ftw_{\ktw}
  - \ftw_{\ktw} \ftw_{\ptw} \Big)
 \nonumber \\ && + 4\pi \int_0^\infty d\ptw d\ktw \;
  \gammatw^g_{gg}(\ptw+\ktw;\ktw,\ptw)
 \Big( \ftw_\ptw \ftw_{\ktw} - \ftw_{\ptw+\ktw} \ftw_{\ktw}
  - \ftw_{\ptw+\ktw} \ftw_{\ptw} \Big) \,,
\eea
which makes it clear that the first term is minus half the second term.
The rate of particle number destruction is therefore
\be
4\pi \int_{\ptw > \ktw} d\ptw d\ktw \;
  \gammatw^g_{gg}(\ptw+\ktw;\ktw,\ptw) \:
  \ftw_{\ptw{+}\ktw} \ftw_\ptw \ftw_\ktw
 \Big( \ftw^{-1}_{\ptw+\ktw} - \ftw^{-1}_{\ptw}
  - \ftw^{-1}_{\ktw} \Big)
 = \frac{\ntw}{7} \,,
\label{number_change}
\ee
where we used \Eq{partnum1} and \Eq{partnumL} to equate the integral to the
particle number.  Note that if $\ftw_\ptw$ is steeper than $\ptw^{-1}$
at all $\ptw$, then the lefthand side is everywhere positive.

Since the particle number is finite, the lefthand side of
\Eq{number_change} must also be finite.  The danger is of a divergence
at small $\ktw$.  In this limit $\gammatw^g_{gg}$ behaves as
\be
\lim_{\ktw \ll 1,\ptw} \gammatw^g_{gg}(\ptw+\ktw;\ptw,\ktw)
\propto \frac{1}{\ktw} \quad \mbox{and independent of $\ptw$.}
\ee
(This fact is familiar from the physics of initial state radiation,
where it gives rise to the log soft divergence in the total emission
rate.)  The product of statistical functions must therefore remain finite
in this limit.  If $\ftw_\ktw$ grows faster than $\ktw^{-1}$ in the
infrared, then $\ftw^{-1}_\ktw$ falls faster than linearly and can be
neglected.  Then approximating
$\ftw^{-1}_{\ptw+\ktw} - \ftw^{-1}_{\ptw} \simeq
\ktw \,d\ftw^{-1}_\ptw / d\ptw > 0$, we find that the integral is small
$\ktw$ divergent.  Therefore $\ftw_\ktw$ cannot grow faster than
$\ktw^{-1}$ in the infrared, in order for the particle destruction rate to
remain finite, within the kinetic description we have followed.

Note that, since $\mDtw$ is finite and the quoted behavior for
$\gammatw$ is only valid for $\ktw \geq \mDtw$, this argument is
only rigorous if $\mDtw \ll 1$, meaning at late times.  Nevertheless,
what it shows is that, {\sl at sufficiently late times and deep enough
  in the infrared,} the behavior
of the occupancy must scale as $f \propto p^{-1}$, not a steeper power
such as $p^{-4/3}$.  This is in contrast to what one might guess based
on certain cascade arguments \cite{Berges}.

\section{Solving the Boltzmann equation}
\label{secSolving}

We have developed two methods for solving the Boltzmann equation, a
variational method which is specialized to the problem at hand and a
time-domain, momentum-discretization approach which should have wider
applications.  We will present each approach in turn.

\subsection{Variational formulation}

\Eq{Boltzmann2} admits a one parameter family of solutions corresponding
to the arbitrary initial value of the energy density. Defining
\be
\epsilon[\ftw] = \int d\ptw ~\ptw^3 ~\ftw_{\ptw}
\label{constraint}
\ee
and recalling our definitions, \Eq{Qdef}, \Eq{ptw} and \Eq{ftw},
we are seeking the unique function $\ftw(\ptw)$ which satisfies
\Eq{Boltzmann2} with $\epsilon[\ftw] = 1$. We will make the problem
variational by specifying an action $\Gamma[\ftw]$ which reaches its
extremum when \Eq{Boltzmann2} and the condition $\epsilon[\ftw] = 1$ are
satisfied. Integrating over the sphere,
\be
\int \ptw^2 \frac{d\Omega}{(2\pi)^2} = \frac{\ptw^2}{\pi}
\ee
we will now define the operators
\begin{eqnarray}
L[\ftw_{\ptw}] &=& \frac{\ptw^2}{7\pi} \Big( 4\ftw_{\ptw} + \ptw \ftw'_{\ptw}\Big ) \\
C[\ftw_{\ptw}] = C_{\onetwo}[\ftw_{\ptw}] + C_{\twotwo}[\ftw_{\ptw}]
 & = & \frac{\ptw^2}{\pi}\Big (
  \tcalC_{\twotwo}[\ftw_{\ptw}] + \tcalC_{\onetwo}[\ftw_{\ptw}] \Big )
\label{newC}
\end{eqnarray}
and choose the action to be
\be
\Gamma[\ftw] =
\lambda \Big ( \epsilon[\ftw] - 1\Big)^2
+ \int \frac{d\ptw}{2\pi} \ptw^{2\alpha} \Big (
L[\ftw_{\ptw}] - C[\ftw_{\ptw}]\Big )^2 \,.
\label{gamma}
\ee
For any choice of $\alpha$ and $\lambda > 0$ this action is nonnegative
definite, but it equals zero where \Eq{Boltzmann2} is satisfied;
therefore it is a good starting point for a variational solution.

There remain technical issues, both in the computation of
$\gamma^g_{gg}$ and in the handling of the multiple integrals involved
in $C_{\twotwo}$.  We postpone these to Appendix \ref{secM12}
and Appendix \ref{secM22} respectively.

Concerning the inclusion of a screening mass, it was mentioned in the
previous section that its presence would break the otherwise exact
scaling law. To proceed onwards, we will simply fix the value of
$\mDtw$, which is a reasonable approximation since the actual dependence
on $\mDtw$ is weak.

Since $\ftw(\ptw)$ has an infinite number of degrees of freedom, we must
make some simplifications in order to seek an extremum.  We choose to
extremize $\ftw(\ptw)$ over some flexible but finite-parameter
\Ansatz; specifically we will consider
\be
\ftw(\ptw) = A \Big ( \frac{\ptw}{\omega_1} \Big )^{g_1(\ptw)}
 e^{-\beta\ptw^{g_2(\ptw)}\frac{\ptw^\gamma}
  {\ptw^\gamma+\omega_2^\gamma}}
\label{fansatz}
\ee
where $g_1$ and $g_2$ are rational functions of the form
\be
g_i (\ptw) = \frac{a_{i,N_i} (\ptw / \lambda_i)^{N_i} + ... + a_{i,0}}{b_{i,N_i} (\ptw / \lambda_i)^{N_i} + ... + 1} .
\ee
The variational coefficients are
$c_i \in \{A,\omega_1,\beta,\gamma,\omega_2,a_{1,0}... \}$. Physically
$\beta$ and $g_2$ control the UV behavior, while $g_1$ primarily
controls the IR behavior.  The extremal
value within this \Ansatz\ is the choice of $c_i$ such that
\be
\label{variation}
\frac{\partial}{\partial c_i} \Gamma[\ftw_{\ptw}] = 0 .
\ee
In the limit where the \Ansatz\ is described by an infinite
number of parameters, this equation in principle becomes exact. The
extremal point of $\Gamma$ can be located iteratively by a
numerical implementation of non-linear conjugate gradient descent, and
the failure to satisfy \Eq{Boltzmann2} exactly can be assessed by
plotting $L[\ftw_{\ptw}]$ and $C[\ftw_{\ptw}]$ as functions of $\ptw$ and seeing with
what accuracy they cancel.

\subsection{Discrete-momentum method}

The second implementation of the Boltzmann equation we have used
involves the direct time evolution of a momentum-discretized version
of \Eq{lhs=rhs} (naturally including both $\tcalC_{2\leftrightarrow 2}$
and $\tcalC_{1\leftrightarrow 2}$).  We do so by introducing a discrete
sample of points $\ptw_i$ and tracking the number density of particles
with momentum near $\ptw_i$, $\ntw_i$.  Specifically, a
continuous distribution $\ftw(p)$ is converted into the discrete
$\ntw_i$ via
\be
\ntw_i \equiv \int \frac{d^3 \ptw}{(2\pi)^3} \ftw(\ptw)
w_i(\ptw) \,, \qquad
w_i(\ptw) \equiv \left\{ \begin{array}{ll}
\frac{\ptw-\ptw_{i-1}}{\ptw_i-\ptw_{i-1}} \,, & \ptw_{i-1}<\ptw<\ptw_i \\
\frac{\ptw_{i+1}-\ptw}{\ptw_{i+1}-\ptw_i} \,, & \ptw_i<\ptw<\ptw_{i+1} \\
0 & \ptw<\ptw_{i-1} \mbox{ or } \ptw>\ptw_{i+1}\,. \\
\end{array} \right.
\ee
Here the ``wedge'' function $w_i(\ptw)$ rises linearly from $0$ at
$\ptw_{i-1}$ to 1 at $\ptw_i$ and then falls to zero linearly at
$\ptw_{i+1}$, so $\sum_i w_i(\ptw) = 1$ for any $\ptw$ within the range
considered.
The points $\ptw_i$ need not be evenly spaced and in practice it is best to
space them more tightly where $\ftw$ shows stronger variation.
In terms of the $\ntw_i$, the particle number and energy densities are
\be
\ntw = \sum_i \ntw_i \,, \qquad
\epsilon = \sum_i \ptw_i \ntw_i \,.
\ee

The time evolution of $\ntw_i$ is determined by integrating
\Eq{lhs=rhs} over $\ptw$,
\be
\label{discreteBoltzmann}
t \partial_t \ntw_i = \frac{1}{7} \left( 4 \ntw_i
  + \ptw \frac{d\ntw}{d\ptw} \right)
- \int \frac{d^3 \ptw}{(2\pi)^3} w_i(\ptw) (\tcalC_{1\leftrightarrow 2}(\ptw)
+ \tcalC_{2\leftrightarrow 2}(\ptw) ) \,.
\ee
The equation is evolved until it converges, yielding the scaling
solution for $\ntw_i$.

Using \Eq{C2x2tw}, the collision term is
\bea
\label{C2xagain}
\int \frac{d^3 \ptw}{(2\pi)^3} w_i(\ptw)
\tcalC_\twotwo & = & \frac{1}{8} \int
\frac{d^3 \ptw d^3 \ktw d^3 \ptw' d^3 \ktw'}{(2\pi)^{12}}
\frac{\Abs{\overline{M}_{\ptw \ktw;\ptw'\ktw'}^2}}{2\ptw2 \ktw 2\ptw'2\ktw'} \;
  (2\pi)^4 \delta^4(\ptw{+}\ktw{-}\ptw'{-}\ktw') \times
\\ \nn &&\hspace{1em}
 \ftw_{\ptw} \ftw_{\ktw} \ftw_{\ptw'} \ftw_{\ktw'}
    \Big( \ftw_{\ptw'}^{-1} {+} \ftw_{\ktw'}^{-1}
    {-} \ftw_{\ptw}^{-1} {-} \ftw_{\ktw}^{-1} \Big)
\Big( w_i(\ptw) {+} w_i(\ktw) {-} w_i(\ptw') {-} w_i(\ktw') \Big)
\eea
and similarly for \Eq{C1x2tw}.  In the numerical implementation all
$\partial_t \ntw_i$ are computed simultaneously; the
values of $p,p',k,k'$ are sampled, and each sample point then
contributes to the eight $\partial_t \ntw_i$ for which a $w_i$ function
is nonzero.  This approach identically conserves total energy and
violates particle number by precisely the amount stipulated in
\Eq{number_change} -- in particular the $2\leftrightarrow 2$ process
exactly conserves particle number under this implementation.

So far the implementation we have described is an exact representation
of the original Boltzmann equation.  The implementation becomes
approximative because we must deal with two quantities which are not
strictly well defined in terms of the $\ntw_i$ alone.  The first
is $\ptw \frac{d\ntw}{d\ptw}$, appearing in \Eq{discreteBoltzmann}.
We can fix it uniquely by the requirement that the rescaling of
momentum and occupancy with time, introduced in \Eq{ptw} and
\Eq{ftw}, identically preserves particle number and energy.  This
leads to
\be
 \ptw \frac{d\ntw_i}{d\ptw} \equiv
 -\ntw_i \frac{\ptw_i}{\ptw_i - \ptw_{i-1}}
 +\ntw_{i+1} \frac{\ptw_{i+1}}{\ptw_{i+1} - \ptw_i} \,.
\ee
The other quantity we must deal with is the occupancy $\ftw_\ptw$
appearing in \Eq{C2xagain}.  We interpolate this
from the $n_i$; for $p_i < p < p_{i+1}$ we use
\be
4\pi p^2 \ftw(p) =
\frac{2 \ntw_i}{\ptw_{i+1}-\ptw_{i-1}}
   \frac{\ptw_{i+1}-\ptw}{\ptw_{i+1}-\ptw_i}
+ \frac{2 \ntw_{i+1}}{\ptw_{i+2}-\ptw_i}
   \frac{\ptw-\ptw_i}{\ptw_{i+1}-\ptw_i} \,.
\ee
The need for this inerpolation means that the method is not exact.  However,
discretization errors in this approach should scale as the second power
of the $\ptw_i$ spacing.  Numerically it is not difficult to implement
200 or more points. In the discrete momentum method, we can set $\mDtw$
by hand as in the \Ansatz\ method, or we can determine $\mDtw$ self-consistently
as an integral moment of the distribution as a function of time.

\section{Results and discussion}
\label{results}
\begin{figure}
\centering{\includegraphics[scale=0.75]{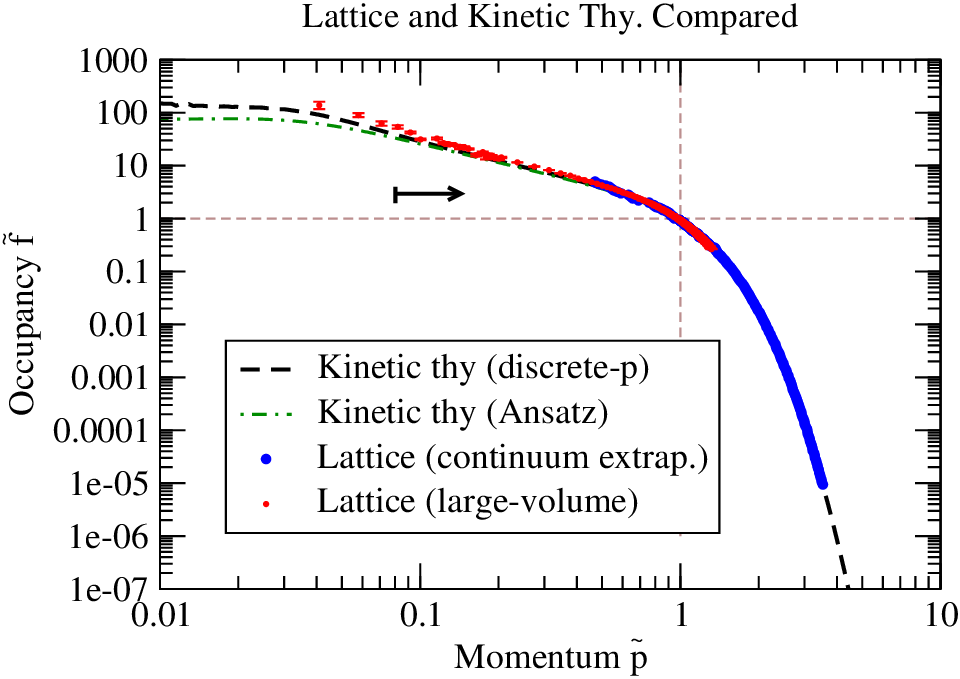}}
\caption{\label{mainresult}
  Scaling solution for the occupancy, when $\mDtw = 0.08$.
  Kinetic theory results use the momentum discretization method
  (dashed black line) and via the \Ansatz\ method (green
 dash-dotted curve).
  Lattice results are shown using a large
  volume (red circles, used in the IR) and a smaller volume
  but with careful extrapolation to the continuum limit (blue
  circles, used in the UV). The beginning of the arrow marks the scale $\mDtw$ above which the lattice and kinetic theory results should agree.}
\end{figure}

As mentioned before, the cascade to the UV does not quite achieve a
scaling solution, because the Debye scale evolves relative to the
characteristic momentum: $\mDtw \sim (Qt)^{-2/7}$.  Therefore we must
fix a value of $\mDtw$ and determine the scaling solution at that
epoch.  Figure \ref{mainresult} shows the scaling solution we find
when $\mDtw = 0.08$ (corresponding to time $Qt=2000$).  The figure shows the results using kinetic
theory solved via the momentum discretization method and via the
\Ansatz\ method.  The curves are nearly identical, except far
in the infrared, $\ptw < 0.1$, where the \Ansatz\ method loses
resolution and where kinetic theory no longer accurately describes
the full (hard-loop) dynamics.

The figure also compares the kinetic theory results with the direct
determination of the occupancies, established by solving the classical
theory on the lattice.  The lattice data are also evaluated at time
$(Qt) = 2000$, when the occupancies self-consistently return a value
$\mDtw \simeq 0.08$. In the infrared it is important to use a large
volume and high statistics, so we have averaged our results over 6
independent evolutions with $(Qa)=0.2$ and $(QL)=51.2$.  In the
ultraviolet it is important to extrapolate carefully to the
small lattice-spacing limit, so we have extrapolated over three
spacings down to $(Qa)=0.1$, with half the box length (the results
remain unchanged if we halve the box size again).  These results
are shown in Figure \ref{mainresult} as red and blue circles, respectively.  Without the continuum extrapolation the
UV tail would not fit the kinetic theory result.  All lattice
data are based on Coulomb gauge-fixed transverse electric field
correlators, using dispersion corrected for plasma frequency
effects as described in Ref.~\cite{KM3}.

The figure shows clearly that kinetic theory provides an excellent
description of the lattice results for the scaling solution of the UV cascade,
except in the infrared, $\ptw < 0.1$.
Note however that the kinetic theory we have used is only strictly valid in
the momentum region $p \gg \mD$, marked by the black arrow in the figure.  In particular, when computing the
collision operator $\calC_{\twotwo}$ we have treated the external states
with massless dispersion and made Hard Thermal Loop approximations
in the matrix elements (and the approximations described in
Appendix \ref{secM22}) which are not reliable in this regime.  In
addition we have continued to treat the $\calC_{\onetwo}$ splitting
process in a collinear expansion which also loses its reliability
for $p \lsim \mD$.  We believe that one could in principle improve
kinetic theory such that it incorporates these effects in the IR, but
to our knowledge this has not been done.

\begin{figure}
\centering{\includegraphics[scale=0.65]{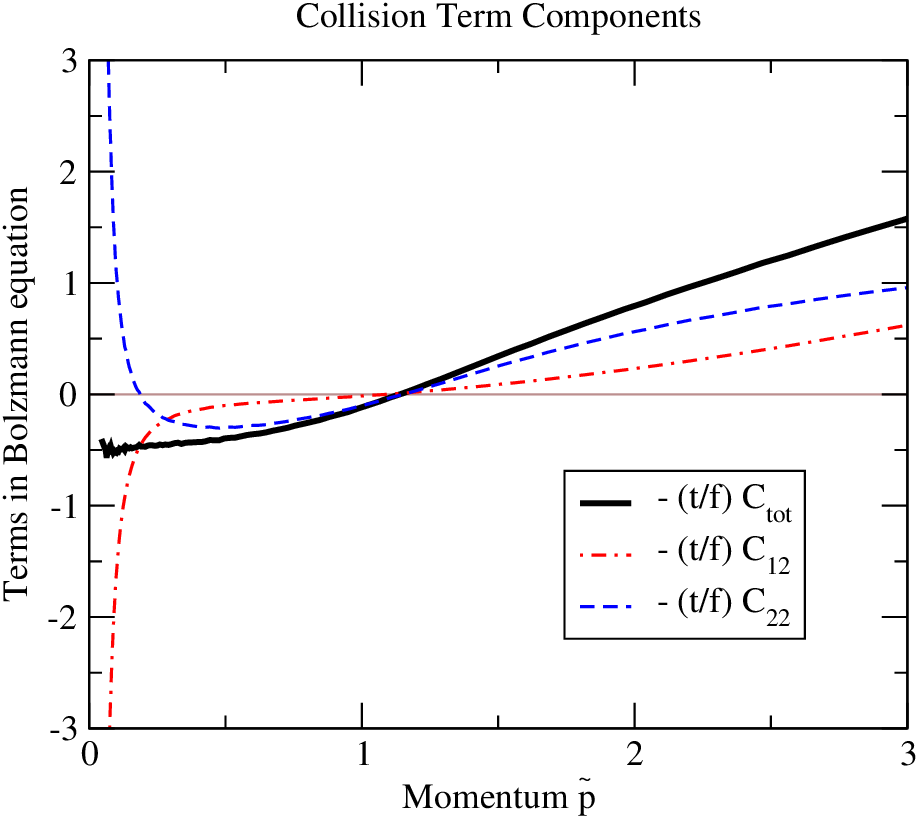}}
\caption{\label{Ccompare}
  Components of the Boltzmann equation, expressed as relative
  contributions to the particle number evolution $(t/f) df/dt$,
  computed with the discrete momentum method with $\mDtw=0.08$.
  $\onetwo$ processes (red dash-dot), $\twotwo$ processes
  (blue dashes), the sum (black solid).
  }
\end{figure}

One advantage of having a kinetic description of the full problem is
that we can determine what physics is most important in controlling the
evolution of the particle cascade.  To explore this, we compare the
relative sizes of $\calC_{\twotwo}$ and $\calC_{\onetwo}$, as a function
of momentum, in Figure \ref{Ccompare}.
The solid (black) line in the figure shows the total occupancy evolution
$(t/f) df/dt$, which switches from negative below $\ptw=1.1$
(particle number leaving the infrared) to positive above $\ptw=1.1$
(particle number filling into the ultraviolet).  The results for the $\onetwo$ and $\twotwo$ collision processes
are shown in red dash-dot and blue dashed, respectively.  We see that
the $\onetwo$ processes raise particle number in the UV above
$\ptw=1.1$ and remove particle number in the IR, while the $\twotwo$
processes raise particle number both in the UV (energy cascade) and
IR (particle number cascade), while removing particles in the range
$0.2 < \ptw < 1.2$. 

The Figure \ref{Ccompare}, along with an analysis of the details of each collision
term, give us information about what the most important processes
are in each momentum range.  The figure shows that the cascade filling
the UV modes is a mixture of the two collision processes.
Since the $\twotwo$ process is enhanced by $\log(1/\mDtw)$, its
relative importance gradually increases as the occupancy falls with
time.  On the other hand, the deep infrared is controlled by a
competition between the two processes, with a large rate of particles
entering the IR via $\twotwo$ processes and a large removal rate
via $\onetwo$ processes.  The IR occupancy is determined by the
requirement that these rates be in balance.

What are the typical momenta involved in these IR occupancy establishing
processes?  We find that the typical $\twotwo$ process is one involving
a small-exchange-momentum collision between an IR particle and a typical
($\ptw \sim 1$) particle.  The typical $\onetwo$ process involves
a soft particle being absorbed onto a typical ($\ptw \sim 1$) particle.
In particular, when $\ptw \ll 1$, \Eq{C1x2tw} is dominated by
$\ktw \sim 1$, not by $\ktw \sim \ptw$.  This is confirmed by the
numerics.  Therefore the processes which establish the occupancy in
the regime $\ptw \ll 1$ are not dominated by scatterings between
particles of comparable momentum, but involve scatterings between soft
particles and typical particles with $\ptw \sim 1$.  This means that the
conditions for an energy cascade \cite{Berges} are not met.

\begin{figure}
\centering{
\includegraphics[scale=0.45]{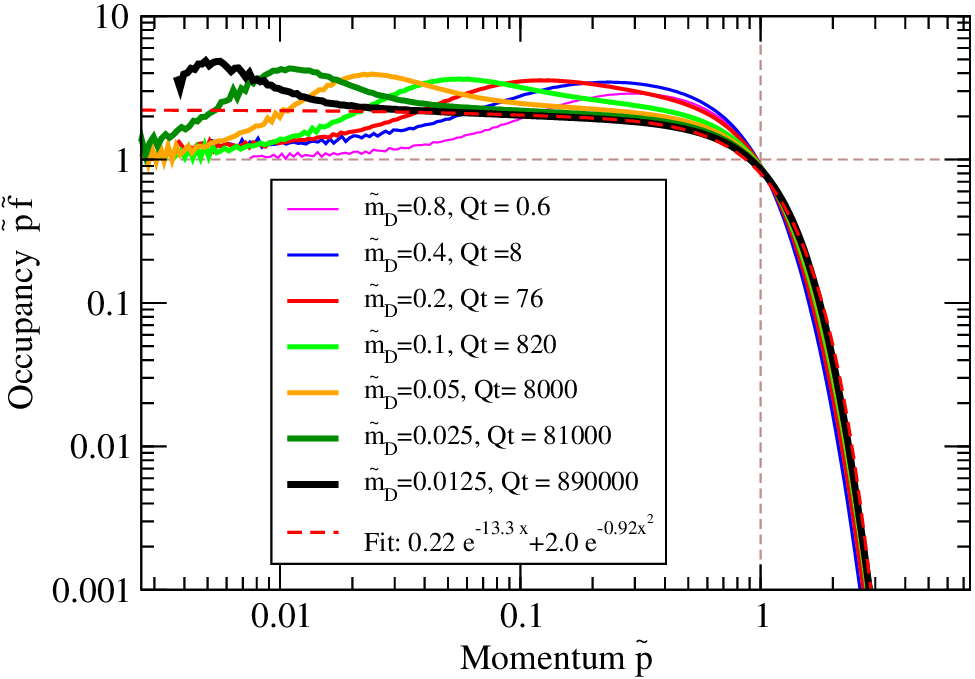}\hspace{0.1cm}
\includegraphics[scale=0.45]{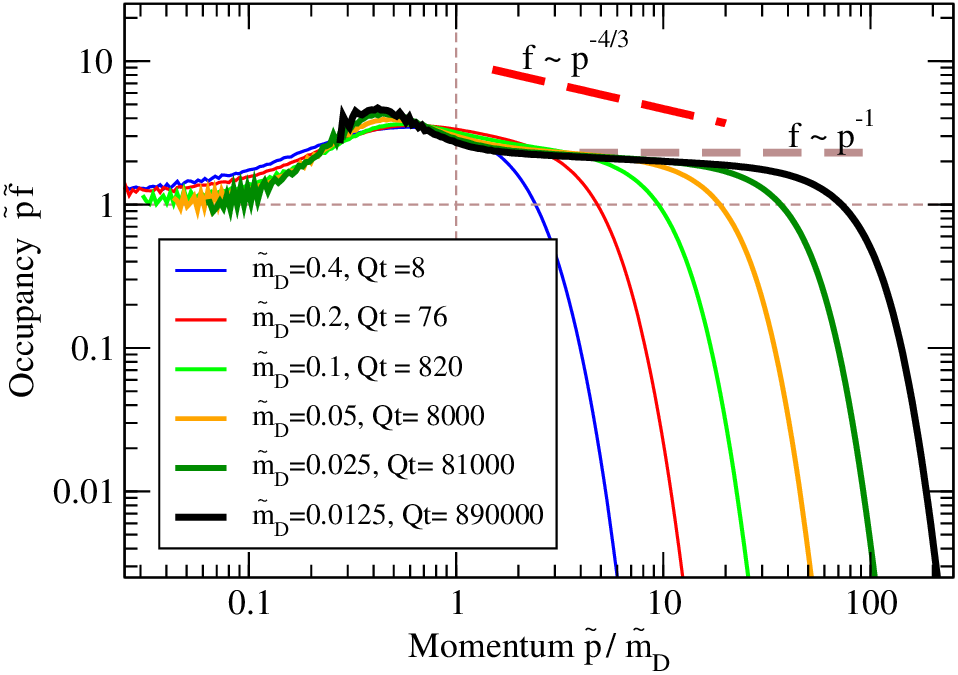}
\caption{\label{md_scaling}
(Left) Evolution of the scaling solution as a function of the screening scale $\mDtw$ computed using the discrete momentum method. The solutions coincide with each other for $\ptw>1$, and exhibit a bump feature at the scale $\mDtw$. The  dashed red line is a fit to the UV part of the spectrum.
(Right) The same as Left, but the $x$-axis is scaled so that the screening scale stays at fixed $x=1$ as a function of time. While at early times when $\mDtw\lesssim 1$, a power-law $\ftw\propto \ptw ^ \alpha$ with $\alpha<-1$ may be seen in the data (in particular $\alpha \sim -4/3$ for $\mDtw \sim 0.1$), at later times when a proper scaling window has emerged, the solution for
$\mDtw < \ptw< 1 $ approaches $\ftw\propto \ptw^{-1}$.
}
}
\end{figure}

How the scaling solution varies with $\mDtw$ is shown in Figure \ref{md_scaling}.
For $\ptw\gtrsim 1$ the solution is highly insensitive to $\mDtw$. However for
large values of $\mDtw$, the $\twotwo$ element is suppressed leading to less
collisions and slightly softer UV tail. For $\ptw \sim \mDtw/2$ the solutions exhibit a bump feature that becomes clearly separated from the UV part of the spectrum for small values of $\mDtw$. For small values of $\mDtw$, where there is proper scale separation between the screening scale and $\pmax$, the solution in the region $\mDtw < \ptw  < 1$ approaches a power law $\ftw \propto 1/\ptw$.
While the peak of the bump is in the region where the kinetic theory does not provide a reliable description, a rise similar to the onset of the bump can be seen also in the lattice data in Figure \ref{mainresult} around $\ptw \gtrsim \mDtw$. For $\mDtw \gtrsim 0.1$ this feature is mixed with the UV tail, and looking at data at these values of $\mDtw$, it is easy to be misled by to data to think that there is
a power law with a higher negative power of $\ptw$ for $\mDtw \lesssim \ptw \lesssim 1$ \cite{KM3,Berges}. We find that the data at $\ptw \gg \mDtw$ is rather well described by a fitting function
\be
\ftw(\ptw) \approx \frac{1}{\ptw}  \left( 0.22 e^{-13.3 \ptw} + 2.0 e^{-0.92\ptw^2}\right),
\ee
also depicted in Figure~\ref{md_scaling}. We expect that in the limit of $\mDtw\rightarrow 0 $ the full scaling solution relaxes to this fit.

\section{Summary}

In summary, we have considered classical Yang-Mills theory with initial
conditions which are statistically isotropic and homogeneous and with
the energy residing in the infrared.  We have
confirmed the existence of a scaling solution for occupancies which
obeys $f\sim(Qt)^{\frac{-4}{7}}$ (endowed with a UV scale
$p_{\text{max}}\sim Q(Qt)^{\frac{1}{7}}$) by directly solving the
Boltzmann equation. The resulting occupancy is in agreement with what
has been observed on the lattice over the range of momenta where the
approximation of massless kinematics is valid; this is a numerical demonstration
of (classical) field-particle duality.

The solution obtained here, however, differs from the lattice findings
in the infrared.  This is expected, as our kinetic description does not
treat infrared excitations as screened.  It does not incorporate the
Landau cut, nor the dispersion relation of plasmons.  These explain the
discrepancy at the scale $\mD$ and below.

The kinetic treatment robustly demonstrates that, at late times when
the scales $\pmax$ and $\mD$ are well separated, there is {\sl not} a
scaling window where occupancies scale as $p^{-4/3}$, extending down to
the scale $\mD$.  Indeed, such
behavior would lead to a divergently large rate of change in particle
number.
However, at intermediate times before a proper scale difference has emerged,
the features at scales $\mD$ and $\pmax$ combine such that result can be easily
misinterpreted --- for a limited range in $p$ --- as a power law with $f\propto p^{-4/3}$.

It would be interesting to
find a way to extend our kinetic treatment to incorporate hard-loop
effects for modes of order the screening scale.  This may be possible,
since the screening scale and the magnetic scale become separated at
large $(Qt)$, so the physics at the $\mD$ scale should be perturbative.

\section*{Acknowledgements}
We thank Jacopo Ghiglieri for useful discussions. 
This work was supported in part by the Canadian National Sciences and
Engineering Research Council (NSERC) and the Institute of Particle
Physics (Canada).

\appendix
\section{$\onetwo$ collision integral}
\label{secM12}

\begin{figure}[h]\centering
\parbox{50mm}{\begin{fmfgraph*}(50,20)
\fmfleft{l3,l2,l1}\fmfright{r4,r3,r2,r1,r0}
\fmflabel{$\theta$}{v2}
\fmf{photon,tension=2}{l1,v1}\fmf{photon,tension=3}{v1,v2}\fmf{photon}{v2,r1}\fmf{photon}{v2,r2}
\fmf{photon,tension=2}{l3,v3}\fmf{photon}{v3,r4}\fmf{photon,label=$q$}{v1,v3}
\end{fmfgraph*}} $\qquad \Longrightarrow \qquad$
\parbox{30mm}{\begin{fmfgraph*}(30,20)
\fmfleft{l3,l2,l1}\fmfright{r5,r4,r3,r2,r1,r0,rn1}\fmfv{label=$\theta$,label.angle=30}{v1}
\fmf{photon,tension=3}{l2,v1}\fmf{photon}{v1,r1}\fmf{photon}{v1,r3}
\end{fmfgraph*}}
\caption{\label{2to3}$2 \rightarrow 3$ processes in the limit where $\theta \sim m_{\text{D}} / p_{\text{max}}$ and $q \sim m_{\text{D}}$ give rise to effective $\onetwo$ splittings.}
\end{figure}
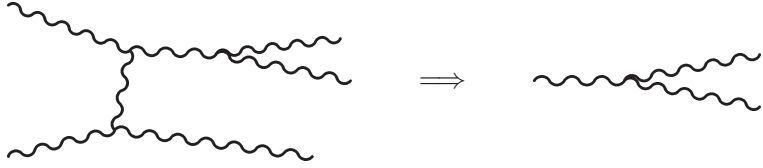

Near collinear $\onetwo$ splitting processes are, strictly speaking,
kinematically not allowed. However, a certain class of higher-order
diagrams, like the $2 \rightarrow 3$ diagram depicted in Fig.~\ref{2to3},
combine and give rise to an effective $\onetwo$ process.

\begin{fmffile}{2to3amplitude}
\end{fmffile}

Consider the contribution of $\twothree$ processes to the collision
term,
\be
\frac{\partial f(p,t)}{\partial t} = \ldots
 - \calC_{\twothree}[f(p,t)] \,.
\ee
As we have seen, if we consider the external states
at fixed energies and angles, the importance of this process declines,
relative to the leading-order one, as $(Qt)^{-4/7}$.  However, the
scattering rate diverges as $\int d^2 q / q^4$, cut off by the soft
scale $\mD$; and even for small $\mD$ the rate for the process with the
extra emitted particle is only down by $\OO(\barf)$ relative to the
process without it, provided that the particle is emitted at an angle
$\theta \lsim q/p$.  Therefore the rate for this process, compared to
the rate for wide-angle $\twotwo$ scattering, is
$\OO(\barf /\mDtwsq) \sim 1$.  But this only occurs in the kinematic
range where $q\sim \mD \ll p$.  In this regime, the statistical factors
for the ``target'' particle (the one which does not split) do not
matter, $(\ftw^{-1}(k) - \ftw^{-1}(k{+}q))\simeq 0$, and we may simplify
the description by only keeping track of the particle which actually
undergoes the splitting.  Higher-order processes
of general form $n\leftrightarrow (n{+}1)$ are also unsuppressed in a
similar kinematic region, so that again only the particle which actually
undergoes splitting need be directly considered.
The contribution to the collision term is
\cite{AMY5}
\bea
\calC_{\onetwo}[f_p] & = & \frac{(2\pi)^3}{p^2\nu_g}
  \int_0^{\frac{p}{2}} dk~\gamma^{g}_{gg}(p;k,p-k)\Big( f_p
  [1+f_k][1+f_{p-k}] - [1+f_p] f_k f_{p-k}\Big) + \nonumber \\
 & & \frac{(2\pi)^3}{p^2\nu_g}
  \int_0^{\infty} dk~\gamma^{g}_{gg}(p+k;p,k)\Big( f_p f_k[1+f_{p+k}] - [1+f_p][1+ f_k] f_{p+k}\Big) \,,
\label{C12}
\eea
where the first term represents the possibility that the particle of
momentum $p$ should split into (or from) two particles of smaller
energy, while the second term represents the production of the particle
of momentum $p$ via the splitting of a higher-energy particle (and its
inverse process).  As before, we make the classical approximation by
replacing
\be
\label{fsimple12}
f_p [1+f_k][1+f_{p-k}] - [1+f_p] f_k f_{p-k}
\simeq f_p f_k f_{p-k} \Big( f_k^{-1} + f_{p-k}^{-1} - f_p^{-1} \Big)\,.
\ee
The splitting kernels $\gamma^g_{gg}$ are effective
matrix elements for these processes.  They were found explicitly in \cite{AMY5}, and are given by
\begin{eqnarray}
\label{gamma_is}
\gamma^g_{gg}(p';p,k) &=& \frac{p'^4 + p^4 + k^4}{p'^3p^3k^3} \mathcal{F}_g(p';p,k)\\
\mathcal{F}_g(p';p,k) &=& \frac{g^2 \nc \nu_g}{4(2\pi)^4}\int \frac{d^2
  h}{(2\pi)^2} \V h\cdot\text{Re}\:\Fg(\V h;p',p,k) ,
\label{calF_is}
\end{eqnarray}
where $\V h = \V p \times \V k$ parametrizes the (parametrically small)
non-collinearity of the external states, and $\Fg$ is the solution of
the following integral equation,
\begin{eqnarray}
2\V h &=& i\delta E(\V h) \Fg(\V h)\nn \\
&&\quad+~\frac{g^2 \nc T_*}{2}\int\frac{d^2\V q}{(2\pi)^2}
 \mathcal{A}(\V q) \big (3\Fg(\V h) - \Fg(\V h{-}p\V q)
                      -  \Fg(\V h{-}k\V q)
                  - \Fg(\V h{+}p'\V q)\big )\qquad
\label{1to2IntEqn}
\end{eqnarray}
with
\begin{eqnarray}
\label{A_is}
\mathcal{A}(\V q) &=& \frac{1}{\V q^2}
                  - \frac{1}{\V q^2 + \mDsq}\, , \\
\label{deltaE_is}
\delta E(\V h) &=& \frac{m^2_\text{D}}{4}\Big ( \frac{1}{p}
                       + \frac{1}{k} - \frac{1}{p'}\Big ) +
                       \frac{\V{h}^2}{2pkp'} \,,
\\
\label{Tstar_is}
T_* & = & \frac{\int d^3 p \: {f}^2(p)}
               {\int d^3 p \: 2 f(p) / p} .
\end{eqnarray}
Now let us determine how these effective $\onetwo$ processes scale with
$g^2 \nc$ and with time.  If we multiply both sides of \Eq{C12}
by $g^2 \nc$, so it describes the evolution of $\barf$ rather than of
$f$, then each factor of $f$ is accompanied by a factor of
$g^2 \nc$ and we can write everything in terms of $\barf$.
Specifically, the collision term is quadratic in $f$, see
\Eq{fsimple12}; but there are two factors of $g^2 \nc$, the one we just
added and the one in the expression for $\mathcal{F}_g$,
\Eq{calF_is}.  The factor $g^2 \nc$ in front of $T_*$ in
\Eq{1to2IntEqn} combines with the definition \Eq{Tstar_is} such
that $g^2 \nc T_* \equiv \barT$ is expressed purely in terms of $\barf$.
The definition of $\mDsq$, \Eq{mD_is}, is also in terms of $\barf$;
$\mDsq \sim \barf \pmax^2$.
Therefore the factors of $g^2 \nc$ all disappear when we work in terms
of $\barf$, just as for $\calC_\twotwo$.  This is the same as the
statement that $\calC_\onetwo$ has a valid classical limit.

Next we must check how everything scales with time, when
$p\sim \pmax \sim Q (Qt)^{1/7}$ and $\barf \sim (Qt)^{-4/7}$.
Because $\barT \sim p \barf \sim \mDsq/p$, the first and
second terms in \Eq{1to2IntEqn} are of comparable size when
$\V h \sim \mD p$, which will be the dominant range for $\V h$
in \Eq{calF_is}.  The magnitude of $\Fg$ is
$\Fg \sim \V h/\delta E \sim p^2/\mD$, so the integral in \Eq{calF_is}
is of order $\Abs{\V h}^3 \Fg \sim \mDsq p^5$ and
$\gamma^g_{gg} \sim \mDsq$.  By \Eq{mD_is},
$\mDsq \sim Q^2 (Qt)^{\frac{-2}7}$, while $p \sim Q (Qt)^{\frac 17}$ and
$\barf \sim (Qt)^{\frac{-4}7}$; so the righthand side of
\Eq{C12} scales as $Q (Qt)^{\frac{-11}{7}}$, exactly the same scaling as
the $\twotwo$ collision term, see the discussion after \Eq{twotwoscales}.

\Eq{1to2IntEqn} is most easily solved by making a transformation to
impact parameter space \cite{Aurenche}; we introduce
\begin{equation}
\V f(\V b) = \int \frac{d^2 q}{(2\pi)^2}~ e^{i\V q \cdot \V b} ~\V f (\V q) ,
\end{equation}
and transform \Eq{1to2IntEqn} into and ODE (choosing units with $\mD=1$)
\begin{eqnarray}
-2i \nabla \delta^{(2)}(\V b) &=& \frac{i}{2p' y(1-y)}\Big (\frac{1-y(1-y)}{2} - \nabla^2 \Big )\V f (\V b) \nn\\
&&~+ \frac{g^2 N_c T}{2}\Big ( D(yb) + D(b - by) + D(b) \Big ) \V f (\V b) .
\end{eqnarray}
In this context, $\V f(\V q)$ is a re-scaling of the original $\Fg$, $\V f(\V q) = \Fg (p' \V q) / p'$. $p'$ is the energy of the incoming gluon, $p' = p+k$, by momentum conservation. Hence, $p = (1-y) p'$, $k = yp'$, where $y$ runs from 0 to 1. The function $D(b)$ is defined by the integral
\begin{eqnarray}
D(b) &=& \lim_{\epsilon\rightarrow 0} \int \frac{d^2 q}{(2\pi)^2} \frac{1}{(q^2+\epsilon^2)(q^2 + 1)} - e^{i\V q \cdot \V b} \frac{1}{(q^2 + \epsilon^2)(q^2+1)}\\
&=& \frac{1}{2\pi}\big (\gamma_E + \log (b/2) + K_0(b)\big ) ;
\end{eqnarray}
$K_0$ is a modified Bessel function. Following the procedure laid out in \cite{Aurenche} and defining $\V f(\V b) = \V b h(b)$, we have
\begin{equation}
\mathcal{F}_g(p';p,k) = \frac{p'^4 g^2 \nc d_A}{2(2\pi)^4}\int \frac{d^2 q}{(2\pi)^2}\V q \cdot \V f(\V q) = \frac{p'^4 g^2 \nc d _A}{2(2\pi)^4} 2~\text{Im}\lim_{b\rightarrow 0^+} h(b) .
\end{equation}
This limit is obtained via the numerical resolution of the ODE; in
practice this quantity only needs to be calculated on some grid of
$(p',y,T_*)$ points. When performing the integral in \Eq{Boltzmann2},
intermediate values of $\gamma^{g}_{gg}$ are then obtained by
interpolation.

The emission and absorption of soft gluons occurs with a divergent rate,
but these processes approximately cancel in \Eq{C12}.  In order to make
the cancellation explicit, we
subdivide $C_{\onetwo}[\ftw_{\ptw}]$ as follows,
\be
C_{\onetwo}[\ftw_{\ptw}] = C^A_{\onetwo}[\ftw_{\ptw}] + C^B_{\onetwo}[\ftw_{\ptw}] + C^C_{\onetwo}[\ftw_{\ptw}] ,
\ee
where (introducing the tilde variables where the momentum scaling has
been incorporated, and absorbing a factor of
$8\pi^2$ into the definition of $\gammatw$)
\begin{eqnarray}
\label{CAIntegral}C^A_{\onetwo}[\ftw_{\ptw}] &=&
  - \int_0^{\frac{\ptw}{2}} d\qtw ~\gammatw^{g}_{gg}(\ptw;\qtw,\ptw-\qtw)\Big( \ftw_{\ptw} \ftw_{\qtw} + \ftw_{\ptw} \ftw_{\ptw-\qtw} - \ftw_{\qtw} \ftw_{\ptw-\qtw}\Big)\\
\label{CBIntegral}C^B_{\onetwo}[\ftw_{\ptw}] &=&
  - \int_{\frac{\ptw}{2}}^{\infty} d\qtw ~\gammatw^{g}_{gg}(\ptw+\qtw;\ptw,\qtw)\Big( \ftw_{\ptw} \ftw_{\qtw} - \ftw_{\ptw} \ftw_{\ptw+\qtw} - \ftw_{\qtw} \ftw_{\ptw+\qtw}\Big)\\
\label{CCIntegral}C^C_{\onetwo}[\ftw_{\ptw}] &=&
  - \int^{\frac{\ptw}{2}}_0 d\qtw ~\gammatw^{g}_{gg}(\ptw+\qtw;\ptw,\qtw)\Big( \ftw_{\ptw} \ftw_{\qtw} - \ftw_{\ptw} \ftw_{\ptw+\qtw} - \ftw_{\qtw} \ftw_{\ptw+\qtw}\Big) .
\end{eqnarray}
The sum $C^{A+C}_{\onetwo} = C^A_{\onetwo} + C^C_{\onetwo}$ is treated
by combining the integrands, which makes the cancellations at small $q$
explicit.

\section{$\twotwo$ collision integral}
\label{secM22}

The $\twotwo$ or elastic collision integral we must consider is
presented in \Eq{C2x2tw}, \Eq{newC}, \Eq{M22}; in addition the matrix
element must be modified by the inclusion of hard loops, as described in
\cite{AMY5}.  We perform the integrations using the parametrization of
the momentum integrals from \cite{AMY5},
\be
\int \frac{d^3 p d^3 k d^3 p' d^3 k'}{(2\pi)^{12} 16 p k p' k'}
(2\pi)^4 \delta^4(p{+}k{-}p'{-}k')
 = \frac{1}{2^{10} \pi^6} \int_0^\infty dq \int_{-q}^q d\omega
 \int_{\frac{q{-}\omega}{2}}^\infty dp
 \int_{\frac{q{+}\omega}{2}}^\infty dk
 \int_0^{2\pi} d\phi \,,
\ee
with $p'=p+\omega$ and $k'=k-\omega$.  However we want to work at fixed
$p$, so we must change the integration order so it is the outermost
integral.  We also find it convenient to bring the $k$ integral outside
the other two and to perform the $q$ integral first;
\begin{eqnarray}
C_{\twotwo}[\ftw_{\ptw}] &=& - \frac{1}{2^{9} \pi^5}
  \int_0^\infty d\ktw \int_{-\ptw}^{\ktw} d\omega
 ~ f_{\ptw} f_{\ktw} f_{\ptw'} f_{\ktw'}
  \Big(f^{-1}_{\ptw'} + f^{-1}_{\ktw'} - f^{-1}_{\ptw} - f^{-1}_{\ktw}\Big)
\times
\nn \\ &&\hspace{1.4cm}
\int_{\Abs{\omega}}^{{\qtw}_{\text{max}}} d\qtw \int_0^{2\pi}d\phi ~
\left\vert \overline{M}^2_{\ptw \ktw;\ptw'\ktw'} \right\vert \,,
\label{C22AMY}
\end{eqnarray}
with
${\qtw}_{\text{max}} = \text{min}\{2\ktw - \omega,2\ptw+\omega\}$.
In terms of these variables, the Mandelstam variables $s$ and $t$
appearing in \Eq{M22} are
\bea
\label{Mandels}
s &=& -\frac{t}{2{\qtw}^2}
  \big( (2\ptw + \omega)(2\ktw - \omega) + {\qtw}^2 \big )
  + \frac{t}{2{\qtw}^2}\sqrt{(4 \ptw {\ptw}' +t)
    (4\ktw {\ktw}'+t)}\cos\phi \,, \\
\label{Mandelt}
t &=& \omega^2 - {\qtw}^2 \, .
\eea

The main challenge associated with $C_{\twotwo}$ is that it contains a 4
dimensional integral that at worst must be entirely computed
numerically.  Furthermore, for small values of $\mDtw$, a large array of
points is required to achieve a reasonable amount of
precision. \Eq{C22AMY} is suggestive, in that the distribution functions
do not participate in the integrals over $\qtw$ and $\phi$. At best, it
may be possible to drastically simplify \Eq{C22AMY} by performing the
$q,\phi$ integrals analytically.  In practice, when we include full hard
loops into \Eq{M22}, the $\phi$ but not the $q$ integral can be done
analytically.  But we will now show that in the current context we can
actually perform the $\qtw$ integral analytically without affecting the
reliability of the result.

Note first that the hard loops only play a role for
$\omega,\qtw \lsim \mDtw$.  There are two possible cases.  Either
one or both of $\ptw,\ktw$ are $\lsim \mDtw$; or
$\ptw,\ktw \gg \mDtw$.  In the former case, the hard-loop treatment of
the matrix element is anyway not reliable, since we do not account for
the modification of the dispersion and spectral weight of the external
states.  In the latter case, we can Taylor expand the statistical
function part of the integrand in small $\omega$.  The lowest nontrivial
term is $\OO(\omega^2)$, corresponding to
drag and momentum diffusion effects.  Higher-order in $\omega$ terms are
insensitive to small $q$.  Any
modification of the matrix element which recovers the same result for
the $\omega^2$ behavior as the full hard-loop treatment does, is equally
accurate.

With this in mind, we make the following substitution in the
$t$-channel denominator:
\be
\label{simpleHTL}
\qtw^2 t \rightarrow t (\qtw^2 + \xi^2 \mDtw^2 ) \,.
\ee
The $u$-channel case is handled by relabeling external states so that it
is the same as the $t$-channel one.
With $s$ and $t$ as defined in \Eq{Mandels} and \Eq{Mandelt},
and the substitution \Eq{simpleHTL}, it is in fact possible to perform
the integrals over $\phi$ and $\qtw$ analytically. The parameter $\xi$
is then fixed by performing the above integrals, and the same integrals
with the full hard-loop self-energy, and choosing $\xi$ so that the
large $\ptw,\ktw$ result, integrated over $\omega^2 d\omega$, is the
same:
\be
\mathcal{I}(\xi,\mDtw) = \int_{-\infty}^{\infty}d\omega ~ \omega^2
  \int_{\Abs{\omega}}^{\infty} d\qtw \int_0^{2\pi}d\phi ~
\left( \Abs{ \overline{M}^2_{\xi,\mDtw}
        }^{\text{Approx HTL}}_{\ktw,\ptw \, \gg \qtw}
   -  \Abs{\overline{M}^2_{\mDtw}
             }^{\text{Exact HTL}}_{\ktw,\ptw \, \gg \qtw} \right) .
\ee
We will therefore obtain the same (integrated) behavior for the
regime $\ptw,\ktw \gg \mDtw$, which is sufficient to ensure that the
approach is as accurate as the full hard-loop approach (described and
advocated in \cite{AMY5}) within the current context.
Numerically we find $\xi = 0.83$, which corresponds well with a ``sum
rule'' value%
\footnote{Recently obtained by Jacopo Ghiglieri, private communication}
of $\xi=e^{5/6}/\sqrt{8}$.

\end{document}